\documentclass{article}[12pt, a4paper]

\usepackage{arxiv}
\usepackage{url}
\usepackage{hyperref}
\usepackage{listings}
\usepackage{xcolor}  % For color support
\usepackage{enumitem}
\usepackage{booktabs}
\usepackage{graphicx}

\definecolor{darkblue}{rgb}{0.0, 0.0, 0.5} % Define a custom dark blue color

\lstdefinelanguage{JSON}{
    basicstyle=\small\ttfamily,
    keywords={true,false,null},
    sensitive=true,
    comment=[l]{//},
    string=[b]",
    morestring=[b]',
    morecomment=[s]{/*}{*/},
    morecomment=[l]{//},
}

\lstdefinelanguage{XPath}{
    keywords={doc, normalize-space, qualify-url},
    sensitive=false,
    comment=[l]{//},
    string=[b]{"},
    morecomment=[s]{/*}{*/}, % Multi-line comments
    morestring=[b]' % Support for single quotes
}

\lstdefinelanguage{HTML5}{
    keywords={<!DOCTYPE, html, head, title, body, h1, h2, h3, p, a, div, span, img, form, input, script, style, link},
    sensitive=false,
    comment=[l]{<!--},
    morecomment=[s]{<!--}{-->},
    string=[b]" % Defines string delimiters
}

\lstdefinestyle{BashStyle}{
    language=Bash,
    frame=lines,
    backgroundcolor=\color{white},
    basicstyle=\small\ttfamily,
    keywordstyle=\color{blue},   % Customize keyword color for Bash
    stringstyle=\color{red},     % Customize string color for Bash
    commentstyle=\color{darkblue},  % Customize comment color for Bash
}

\lstdefinelanguage{YAML}{
  keywords={@url, @xpath, @fields},
  sensitive=false,
  comment=[l]{\#},
  morestring=[b]",
}

\title{Dr Web: a modern, query-based web data retrieval  engine}

\author{
    Ylli Prifti \\ Birkbeck, University of London \\ Email: y.prifti@bbk.ac.uk
    \and
    Alessandro Provetti \\ Birkbeck, University of London \\ Email: a.provetti@bbk.ac.uk
    \and
    Pasquale de Meo \\ DICAM, University of Messina\\ Email: pdemeo@unime.it
}

\begin{document}
\maketitle

\begin{abstract}
This article introduces the Data Retrieval Web Engine (also referred to as \textit{doctor web}), a flexible and modular tool for extracting structured data from web pages using a simple query language. We discuss the engineering challenges addressed during its development, such as dynamic content handling and messy data extraction. Furthermore, we cover the steps for making the DR Web Engine public, highlighting its open source potential.
\end{abstract}

%------------------------------------------
\section{Introduction and Related Work}
The World Wide Web reached 1 billion registered websites in 2014 and was fast approaching 2 bullion by December 2021%
\footnote{ca. 1.92B as of December 2021 according to \href{https://www.internetlivestats.com/}{www.internetlivestats.com} }. 
As of the end of 2024 there are 5.5 billion users of the internet, most generating content each day%
\footnote{Source: \href{https://www.itu.int/en/ITU-D/Statistics/Pages/stat/default.aspx}{www.itu.int}}. 
Most estimations of the internet size are usually based on the number of indexed pages on the leading search engines. 
Counters are generally in the form of users, number of pages, number of websites, number of tweets, etc.
In reality, it is a non-trivial quest to determine the memory size of the internet. 
The situation becomes more challenging if we consider the deep web, which is usually estimated to be much larger than the visible web.

Nevertheless, the indeterministic characteristic of the memory size of the internet, the number is bound to be large and ever-growing. 
The amount of data presents unprecedented opportunities for data mining and information extraction from the web. 
This has proven to be true given the number of scientific papers and research based on data from the web.

However, the web is unstructured. Previous tentatives to apply a machine-readable structure \cite{berners2001semantic} to the web have failed to become large-scale standards. 
As such, in the modern days, data on the web are either made available by their owners in the form of temporal datasets or extracted using crawlers and scraper that leverage existing APIs%
\footnote{Short for \textit{`application programming interface.'}} %
or public web pages. 

Large mainstream online social networks and often well-established social media sites offer access to their data via APIs. 
Methods for leveraging API access for research purposes can usually be found in literature~\cite{Marengo2020MiningAssessment,Lomborg2014Using,Oeldorf-Hirsch2020WhoFacebook,MaiMonicaandLeung2020BigTwitter}.

Even though data mining via APIs is the easiest way to access structured data directly, it comes with challenges and issues. 
For example, Pfeffer et al. \cite{pfeffer2018tampering} showed how to tamper with twitter's sample API. 
Online social networks are massive, and APIs only allow for sampled or local%
\footnote{The reach of the starting node usually determines locality, for example, friends on Facebook, or a sampled search of tweets with a certain hashtag.} %
data access. 
The locality is determined by the point of view of a particular profile, group, tweet, or hashtag. 
Hence, even when APIs allow access to structured data, we often find in literature alternative approaches. 
For example, to build a holistic view of the data on Facebook \cite{Gjoka2010WalkingOSNs} while Catanese et al. crawled 12.5M profiles on Facebook with a Breadth-First-Search crawler \cite{Catanese2011CrawlingPurposes}. 

Web Scraping is widely used in the business world and for scientific purposes. 
Sirisuriya categorised the different techniques in the following groups \cite{sirisuriya2015comparative}:

    \begin{enumerate}\label{web-scr-techniques}
        \item Traditional copy and paste;
        
        \item Text grabbing and regular expression;
        
        \item Hypertext Transfer Protocol (HTTP) Programming;
        
        \item Hyper Text Markup Language (HTML) Parsing;
        
        \item Document Object Model (DOM)Parsing;
        
        \item Web Scraping Software;
        
        \item Vertical aggregation platforms;
        
        \item Semantic annotation recognising, and
        \item Computer vision webpage analysers.
    \end{enumerate}

In a more recent state-of-the-art analysis on web scraping, Sarr et al. \cite{Diouf2019WebApplication} apply a different categorisation based on the `approach:'

\begin{enumerate}
    \item Mimicry; 
    \item Weight Measurement;   
    \item Differential, and 
    \item Machine Learning. 
\end{enumerate}

The considerations above need also be seen from another dimension. 
The web tends to be divided into three categories based on its reachability.  
When we discuss the web, we commonly refer to the "Surface web" that tends to be reachable from traditional, mainstream web engines. However, an even bigger and more information qualitative\cite{Bergman2001} part of the web is the `Deep Web.' 
The Deep Web is usually hidden behind passwords, not linked to or many links deep that are difficult to reach with the traditional approaches of web crawling. 
Dedicated methods are found in the literature that addresses Deep Web data extraction. 
Of particular  interest  for our research is the work of Gottlob et al. \cite{furche2013oxpath} on OXPath - an XPath extension for web crawling and scraping that is particularly successful in extracting data from the deep web.

OXPath uses a CLI and queries written in the extended XPath language (i.e. oxpath) to extract semi-structured data from the web. 
The tool was used as one of the two query engines in a large scale distributed system for data extraction discussed in Prifti's doctoral research~\cite{prifti2023emergence} (Chapter 5 and 6). 

%-------------------------------------
\subsection{The queryable web}

The concept of \textit{queryable} is often tightly coupled with structure. 
Due to the success of the `Structured Query Language' \cite{hursch1988sql}, efforts to make the web queryable often took the form of SQL extensions \cite{spertus2000squeal}. 
However, that doesn't overcome the problem of "structure". The web is unstructured, with patterns emerging between sites of similar categories. 

The Semantic Web \cite{berners2001semantic} and the query languages build on The Semantic web standard, such as RDF \cite{lassila1998resource}, are viable solutions to queryability and structure on the web. On the other hand, the Semantic Web has not been adopted as a standard at speed it was initially expected\cite{mcbride2002four}. 
Much of the web is not structured according to the Semantic Web standard. 

Other approaches are derived from the markup nature of the web and the output as seen by the end users. 
Web users visually consume browser interpretation of HTML content incorporating other media, styling and JavaScript. 
Regular expressions, Document Object Model and HTML Parsing, are methods \cite{sirisuriya2015comparative} that use the markup nature for searching, parsing and extracting content from web pages. 
In a comparison work \cite{Gunawan2019/03} between Document Object Model (DOM), Regular Expressions (RegEx) and XPath\cite{clark1999xml}, RegEx and XPath have similar performances in memory usage and speed of extraction and, unlike DOM, can be used as queries, be external, decoupled and instrumental to the web scraping implementation. 
Between the two, XPath is a query language working well with markup constructs whilst Regular Expressions can easily grow in complexity\cite{Ehrenfeucht1974} because you have to manage the flexibility and different styles of writing HTML (with multiple spaces, double quotes, single quotes, no quotes, in one line, in multi-lines, with inner data, without inner data).

The superiority of XPath to Regular Expressions (and DOM, to the extent that DOM is an unsuitable choice) is further confirmed by the fact that XPath is supported by most modern developer tools for searching (e.g. Chrome dev tools, Firefox dev tools) and modern testing engines like Selenium and Playwright.

\begin{itemize}
    \item \textbf{OXPath - XPath Queryability}
\end{itemize}

The efficiency of building an XPath-based query engine for extracting data from the web has already been shown with OXPath \cite{Sellers2011}. 

OXPath is open source, and has an advanced command line interface\footnote{Please see \href{https://sourceforge.net/projects/oxpath/files/oxpath-cli/1.0.1/}{sourceforge.net/projects/oxpath/files/oxpath-cli/1.0.1/} \label{footnote:oxpath-cli}} that can be used as a web query engine.

Research on OXPath and its implementation has been carried out at Oxford University at the beginning of the last century. 
The OXPath implementation is also used commercially and quoting from the open-source GitHub repository%
\footnote{Please see \href{https://github.com/oxpath/oxpath}{github.com/oxpath/oxpath}.} 

\begin{quote}
    \textit{Meltwater uses OXPath to extract millions of documents from 100'000s of sources daily.} 
\end{quote}

OXPath has fundamental characteristics that make it a preferred choice to other similar web scraping tools. 
These have been discussed largely in \cite{furche2013oxpath}.

\begin{enumerate}[label=(\alph*)]
    \item The OXPath Language construct is a superset of XPath. 
    XPath is an established query language for markup constructs such as XML or HTML; 
    
    \item OXPath supports "Kleene Star" navigation and follows up actions with additional constructs for termination conditions;
    
    \item Extraction markers are embedded in the definition and transparent to the overall construct;
    
    \item Support for actions and user interaction simulation, and 
    
    \item Full support of the XPath node navigation functions. 
\end{enumerate} 

Whilst the OXPath implementation%
\footnote{Please see again \href{https://github.com/oxpath/oxpath}{github.com/oxpath/oxpath}.} %
has been used in automatic full-site extraction~\cite{furche2014diadem, prifti2023emergence}, redundancy driven data extraction~\cite{guo2019red} and browserless web data extraction~\cite{fayzrakhmanov2018browserless}, it is starting to show its limitations (for example, it being tightly coupled to a specific browser version), it also provides further opportunities for improving on scalability, sustainability and performance. 

%----------------------------------------
\paragraph{Data Retrieval Web Engine - JSON Queryability}

During our research and building OXPath queries, we found ourselves in the following situations where OXPath wasn't the best choice. 

\begin{enumerate}[label=(\alph*)]
    \item \textbf{Simpler request} - The navigability of the resulting query didn't require actions and user interaction or browser rendering, but rather simpler requests and link follows would fulfil the needs;
    
    \item \textbf{Pre-Actions and Batch Requests} - The need to perform pre-actions (such as logging in) and then retain user cookies for a batch of links provided in input.
    
    \item \textbf{Support for Modern Browser} - The latest OXPath CLI was built in 2017. 
    It has embedded gecko drivers%
    \footnote{See \href{https://github.com/mozilla/geckodriver}{github.com/mozilla/geckodriver}. }, Selenium and Firefox versions that are, at the time of writing, over five years old. More modern web technologies sometimes have unexpected or unsupported behaviour. 
    
    \item \textbf{Closer to output format} - OXPath was built with XML in mind, and its primary output is XML format. 
    We have seen in the last decade a shift in web technologies from mainly XML-based (SOAP, Web Services, XHTML) to JSON-based (REST APIs, JSON+LD, GraphQL). 
    There might be a need for a query definition that is closer to the output. 
    Additionally, JSON has been successfully used as API query definition, for example, for API queryability in GraphQL\cite{hartig2018semantics}
\end{enumerate}

With this in mind, we set on a journey to build an open-source python package that supports JSON queries, uses the latest gecko driver, browsers and overall is build on top of more modern technologies. 
Our goal is to create an open source tool that can grow to make the whole web queryable. 

%-----------------------------------
\section{The project}

This is the third iteration of building the data retrieval web engine (aka \textit{dr-web-engine} or \textit{Doctor web}). The previous two versions were published as pre-build version in the python packages repository \href{https://pypi.org/}{pypi.org}%
\footnote{The last publication of the pre-release version published on the September 6 2020: \href{https://pypi.org/project/dr-web-engine/0.3.2.2b0/}{pypi.org/project/dr-web-engine/0.3.2.2b0/}.}. 

The source code is published on the GitHub repository \href{https://github.com/starlitlog/dr-web-engine}{github.com/starlitlog/dr-web-engine} and is open for community contributions. The package can be installed from the source code, as explained in the repository README file, or more conveniently, can be installed from the python package repository `\textit{pip install dr-web-engine}`%
\footnote{The head version on the pypi.org repository \href{https://pypi.org/project/dr-web-engine/}{pypi.org/project/dr-web-engine/}.}.

Alternatively, a docker image containing the latest version of the python package, is published in the public Docker Hub repository \href{https://hub.docker.com/r/starlitlog/dr-web-engine}{hub.docker.com/r/starlitlog/dr-web-engine} the image can be pulled using docker pull `\textit{docker pull starlitlog/dr-web-engine}` and used to run queries as shown in the example below. 

The following listings show an example query (file name 4chan-query.json5), its execution using the installed package or the docker image, and an extract of the result.

\begin{lstlisting}[basicstyle=\small\ttfamily, caption=A query example 4chan-query.json5,language=JSON,firstnumber=1, label={lst:4chan_query}]
{
  "@url": "https://boards.4chan.org/pol/catalog",
  "@steps": [
    {
      "@xpath": "//div[contains(@class, 'thread')]",  // Selecting each thread
      "@fields": {
        "title": ".//div[contains(@class, 'teaser')]/text()",  // Extracting thread title
        "link": "./a/@href",  // Link to the thread
        "number_of_posts": ".//div[contains(@class, 'meta')]/text()"  // Number of posts
      }
    }
  ]
}
\end{lstlisting}

\begin{lstlisting}[language=bash, style=BashStyle, caption=Running the query using python package or docker]
    # query execution in verbose mode with headless browser 
    dr-web-engine -q 4chan-query.json5 -o 4chan-data.json -l info --xvfb

    # or alternatively by running the docker image
    docker run --rm -v ~/data:/app starlitlog/dr-web-engine -q 4chan-query.json5 \n
    -o 4chan-data.json -l info --xvfb
    
\end{lstlisting}

\begin{lstlisting}[basicstyle=\small\ttfamily, caption=Extract from the result 4chan-data.json,language=JSON,firstnumber=1]

[
...
  {
    "title": "Hinduphobia is about to become illegal in America...",
    "link": "//boards.4chan.org/pol/thread/497716745",
    "number_of_posts": "R: 82 / I: 14\u25b6"
  },
  {
    "title": "/ptg/ - PRESIDENT TRUMP GENERAL - FIXING THE FARMS EDITION...",
    "link": "//boards.4chan.org/pol/thread/497716315",
    "number_of_posts": "R: 161 / I: 82\u25b6"
  },
  {
    "title": ">Freedom of speech caused the Holocaust This is actually...",
    "link": "//boards.4chan.org/pol/thread/497720014",
    "number_of_posts": "R: 14 / I: 2\u25b6"
  },

...
]

\end{lstlisting}

This reportintentionally avoids too much detail on the technical aspects of the package development. 
For a more in depth reading, we refer the reader to the GitHub repository and connected resources. 
With that in mind, there are some aspects of the package that are worth mentioning because of its unique and diversifying characteristics. 

%------------------------------------------
\subsection{The Query Language for data extraction}
The package support two formats for writing the queries:

\begin{enumerate}
    \item \textbf{JSON5}: Backward compatible with JSON (i.e. JSON is always a valid JSON5), the newer version supports some additional nice features that make it a better fit as a query language. 
    Among others, support from comments in code. 
    
    \item \textbf{YAML}: The less verbose syntax and improved readability makes YAML preferred in some contexts, and we support YAML from the outset queries from the outset. 
\end{enumerate}

JSON5 and YAML are a clear diversion from the OXPath syntax, which was XPath based. Additionally, we decided to diverge the semantics of the query language from OXPath with the intention to simplify writing queries and improve readability. 
While in some cases this might translate in more verbose queries, these are more readable and intuitive to write. 
In the OXPath equivalent query for the 4chan extraction in listing~\ref{lst:xpath_example} it is not immediately clear the structure of the output. 
The equivalent YAML query in listing~\ref{lst:yaml_example} supported by \textit{doctor web} of the same example~\ref{lst:4chan_query} is more readable. 
Additionally, if you remove the query syntax keywords from the JSON5 example in~\ref{lst:4chan_query}, you are left with exactly the structure of the output, creating a clear link between the query and the result.

\begin{lstlisting}[basicstyle=\small\ttfamily, caption=OXPath query for scraping 4chan threads, label={lst:xpath_example}, language=XPath]
doc("http://boards.4chan.org/pol/catalog")
  //div[contains(@class, 'thread')]:<links>[
     .//div[contains(@class, 'teaser')]:<title=normalize-space(.)>
     ./a:<link=qualify-url(@href)>
     .//div[contains(@class, 'meta')]:<number_of_posts=normalize-space(.)>
  ]
\end{lstlisting}

\begin{lstlisting}[basicstyle=\small\ttfamily, caption=YAML equivalent for the 4chan query, label={lst:yaml_example}, language=YAML]
@url: "https://boards.4chan.org/pol/catalog"
@steps:
  - @xpath: "//div[contains(@class, 'thread')]"  # Selecting each thread
    @fields:
      title: ".//div[contains(@class, 'teaser')]/text()"  # Extracting thread title
      link: "./a/@href"  # Link to the thread
      number_of_posts: ".//div[contains(@class, 'meta')]/text()"  # Number of posts
\end{lstlisting}

While it is not the intention of this report to expand on extendability, we want to highlight the simplicity of the language model and the easy with which it can be expanded as shown by the keyword definitions and mapping shown in the mode file~\footnote{\url{https://github.com/starlitlog/dr-web-engine/blob/main/engine/web_engine/models.py}}

For completeness of the query language construct and semantics, listing~\ref{lst:childcare} shows a more complex query that extracts child minder profiles from a search and follows the profile links to extract reviews and other profile information.

\begin{lstlisting}[basicstyle=\small\ttfamily, caption=A more complex query example childcare-query.json5,language=JSON,firstnumber=1, label={lst:childcare}]
{
  "@url": "https://www.childcare.co.uk/search/Babysitters/DA12+1AB",
  "@steps": [
    {
      "@xpath": "//div[contains(@class, 'search-result')]",  // Selecting each search result
      "@fields": {
        "full_name": ".//div[contains(@class, 'items-baseline')]/div[1]/span[1]/text()",
        "ratings": ".//div[contains(@class, 'rating')]/span[1]/text()",
        //"distance": ".//span[contains(@class, 'distance')]/span[2]/normalize-space()",
        "image_url": ".//div[contains(@class, 'profile-image')]//img[1]/@src"
      },
      "@follow": {  // Following the profile link
        "@xpath": ".//div[contains(@class, 'profile-image')]//a[1]/@href",
        "@steps": [
          {
            "@xpath": "//div[contains(@class, 'profile featured')]",  // Extracting profile details
            "@name": "profile",
            "@fields": {
              "bio": ".//h3[text()='About Me']/../p/normalize-space()",
              "experience": ".//h3[text()='My Experience']/../p/normalize-space()"
            }
          },
          {
            "@xpath": "//div[@id='reviews']//div[contains(@class, 'review')]",  // Extracting reviews
            "@name": "reviews",
            "@fields": {
              "reviewer": ".//p[2]//a/text()",
              "reviewer_profile": ".//p[2]//a/@href",
              "rating": ".//div[contains(@class, 'rating')]//img/@alt",
              "comment": ".//p[1]/normalize-space()"
            }
          }
        ]
      }
    }
  ]
}
\end{lstlisting}

\subsection{Extraction engines}

Data retrieval web engine uses Playwright ~\url{https://playwright.dev/python/} as its extraction engine. Playwright has its own embedded web drivers, removing the need to independently manage the drivers. In contrast, OXPath uses Selenium ~\url{https://www.selenium.dev/}, which in turn requires additional web-drivers (e.g. Chromium web drivers can be found here~\url{https://sites.google.com/chromium.org/driver/}). OXPath uses a fixed version of Selenium, which also supports a fixed version for the web drivers. This becomes an important limitation since the web browsing technologies advance at a fast pace (e.g. Firefox had almost 50 releases in 2024 alone \url{https://www.mozilla.org/en-US/firefox/releases/}). It is a non-trivial task to update and support the latest browsers with OXPath while it comes out of the box with \textit{doctor web}. We find the limitation of support for up-to-date browser version the biggest limitation of OXPath and one of the main drivers for writing \textit{doctor web}.   

While Playwright is the main engine for extraction, it is by no means intended to be the only one. In fact, other engines have been considered, including Selenium as a Playwright alternative and httpx~\footnote{\url{https://pypi.org/project/httpx/}} as a fast alternative when rendering the output in a browser is not needed. Doctor web was build with extendability in mind, and the extraction engine is abstracted~\ref{lst:engine_abstraction} away to allow additional engines to be added by community contributions. 

\begin{lstlisting}[basicstyle=\small\ttfamily, caption=Extraction engine abstraction,language=Python,firstnumber=1, label={lst:engine_abstraction}]
from abc import ABC, abstractmethod
from typing import Dict, List, Any


class BrowserClient(ABC):
    """Abstract base class for browser clients."""

    @abstractmethod
    def navigate(self, url: str) -> None:
        """Navigate to a URL."""
        pass

    @abstractmethod
    def query_selector(self, selector: str) -> Any:
        """Query the DOM for an element matching the selector."""
        pass

    @abstractmethod
    def close(self) -> None:
        """Close the browser."""
        pass
\end{lstlisting}

Building doctor web has been a fascinating journey. 
Our approach was to create a robust solution to web scraping that could handle the complexity of modern web pages. 
We incorporated a branching strategy, extensive unit tests, and documentation to encourage open-source contributions.

%-------------------------------------------
\section{Performance and Benchmarking}
In this section, we are going to look at the performance comparison between doctor web and OXPath. 

\subsection{Experiment design}

Our intention is to measure the mean of execution time, CPU usage and memory usage over 100 executions of three extraction queries, respectively designed to be of simple complexity, medium complexity and high complexity. In addition, the same experiment is repeated three times on three different machines intended to represent low performance machines, medium performance and high performance computing power and memory availability.

\begin{table}[htbp]
    \centering
    \caption{Experimental Setup for Execution Time, CPU Usage, and Memory Usage Measurements}
    \small % Change font size for the table
    \begin{tabular}{@{}lll@{}}
        \toprule
        \textbf{Parameter}            & \textbf{Description}                                     & \textbf{Values}                              \\ 
        \midrule
        \textbf{Execution Queries}    & Type of extraction queries                               & Simple, Medium, High                        \\ 
        \midrule
        \textbf{Execution Count}      & Executions each query and each machine                    & 10/query/machine                                        \\ 
         \midrule
        \textbf{Total Execution Count}      & Total number of executions                     & 90                                        \\ 
        \midrule
        \textbf{Performance Categories} & Performance categorization of machines                  & Low, Medium, High                           \\ 
        \midrule
        \textbf{Machines}             & Number of machines used for experiments                  & 3                                           \\ 
        \midrule
        \textbf{Metrics Measured}     & Average metrics recorded during the experiments         & Mean Execution Time, CPU and Memory Usage \\ 
     \midrule
        \textbf{Environment Isolation} & Measures to isolate the experimental environment        & Docker Containers for OXPath and Doctor Web \\ 
        \bottomrule
    \end{tabular}
    \label{tab:experimental_setup}
\end{table}

While it is somehow subjective the qualitative attributes of simple, medium, high for query complexity and similarly the values of low, medium to high for computational power, we explain here some of the characteristics that drive the categorisation. 

\begin{enumerate}
    \item \textit{Simple} A query that accesses a single page and return a limited number of properties (no more than 5), without any complex objects (single values or arrays) with the output file size no more than 0.5M
    \item \textit{Medium} A query that accesses a single page but can return multiple values (more than 5) and has complex objects (a property can be a nested object). The file size can be more than 0.5M
    \item \textit{High} A query that need to access multiple pages to build the output result utilizing Kleene star features, has multiple nested complex objects
\end{enumerate}

Following a similar pattern for computational power: 

\begin{enumerate}
    \item \textit{Low} Utilising a computer that has no more than 4 cores and no more than 8GB of virtual memory available. We utilised a MacBook Air 13-inch 2017 as low configuration. Properties: \textit{CPU: 1.8GHz Intel i5 4 Cores, MEM: 8GB 1600MHz DDR3}
    \item \textit{Medium} Utilising a computer that has no more than 8 cores and no more than 96GB of virtual memory available. We utilised an iMac 2019 as medium configuration. Properties: \textit{CPU: 3GHz Intel i5 6 Cores, MEM: 72GB 2667MHz DDR4}
    \item \textit{High} Utilising a computer that highly outperforms the medium configuration.  We utilised a Lenovo ThinkStation P920 as high configuration. Properties: \textit{CPU: 2.3GHz Intel Xeon Gold 5118 48 Cores, MEM: 128GB 2133MHz DDR4}
\end{enumerate}

Additionally, to partially reduce noise by the high diversity of the systems used for benchmarking, we utilized the same docker image with the same configuration to run the experiment. On all occasions, we configured docker to utilize max 2 CPU cores and 1GB of virtual memory. 

\subsection{Experiment results}

All experiment setup, resulting data and analysis Jupyter notes can be found on GitHub~\footnote{\url{https://github.com/starlitlog/dr-web-engine/tree/main/benchmark}}. We executed the experiment multiple times and consistently found the following: 

\begin{enumerate}
    \item Doctor Web is faster, reducing the execution time by half on average across all different computational specs and query complexity.
    \item Doctor Web requires on average 40\% less CPU usage across all different computational specs and query complexity.
    \item Doctor Web uses about 60\% less memory cross all different computational specs and query complexity.
\end{enumerate}

\begin{figure}
    \centering
    \includegraphics[width=0.6\linewidth]{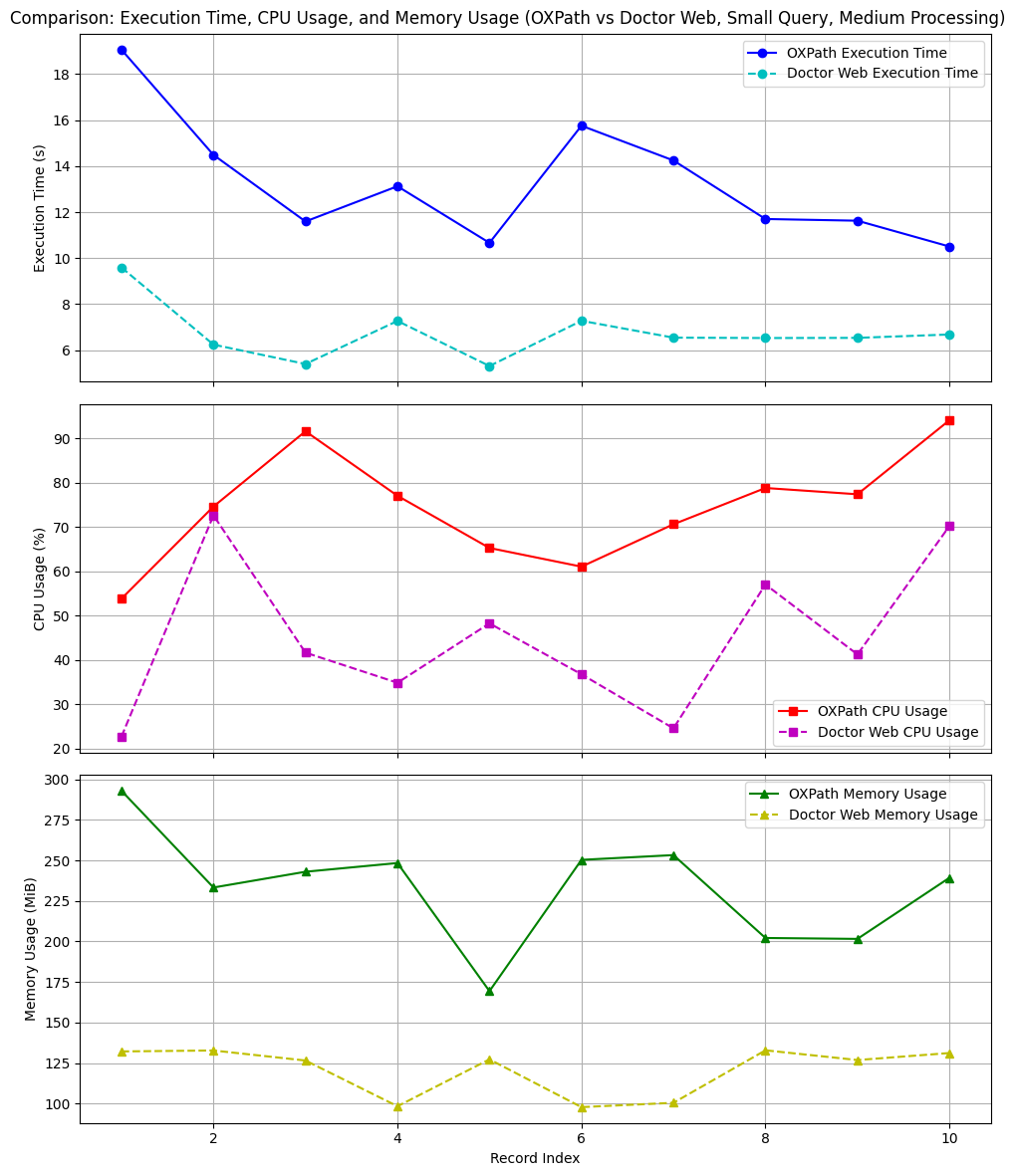}
    \caption{Comparing Doctor Web and OXPath}
    \label{fig:comparison}
\end{figure}

In Figure~\ref{fig:comparison} we show graphically show the results of running the same simple query with OXPath and Doctor Web on a medium powered computational machine.

There are some limits with the results due to the fact that we ran the experiment using docker images and the docker setup would affect the overall efficiently. While the results show a clear picture in terms of execution time, memory and CPU usage, there is a long way ahead before being conclusive due to the fact that there are clear feature gaps between OXPath and Doctor Web with the former being a feature mature engine with many years of usage.

As Doctor Web dynamically grows and the feature gab becomes smaller, the benchmarking experiments will become an important tool to measure if the improvement will resist time and increase in complexity. 

%-------------------------------------------
\section{Conclusion and Future Work}
The development of the DR Web Engine is an ongoing process. We count on the tool growing dynamically by community contribution. We invite contributions and feedback to refine and enhance its capabilities. 

While there are some clear feature gaps with existing established engines like OXPath, we show that Doctor Web has huge growth potential because of its modularity, build for community contribution, reduced complexity for query writing and support for different query formats. Additionally, we show that Doctor Web on average improves efficiency of memory and CPU usage by 40\% and reduced the execution time by half. 

In future work, we plan to tackle further challenges in web scraping and continuously improve the engine's efficiency and reliability.

%-------------------------------------------
\section*{Acknowledgments}
We are grateful to Prof. Georg Gottlob FRS who introduced and motivated us to work on web data extraction.
Thanks to all contributors and users who provided valuable feedback throughout the development process.

%------------------------------------
\bibliographystyle{unsrt}
\bibliography{ylli-thesis-bib.bib}

\begin{thebibliography}{10}

\bibitem{berners2001semantic}
Tim Berners-Lee, James Hendler, and Ora Lassila.
\newblock The semantic web.
\newblock {\em Scientific american}, 284(5):34--43, 2001.

\bibitem{Marengo2020MiningAssessment}
Davide Marengo, Danny Azucar, Claudio Longobardi, and Michele Settanni.
\newblock {Mining Facebook data for Quality of Life assessment}.
\newblock {\em Behaviour {\&} Information Technology}, 0(0):1--11, 2020.

\bibitem{Lomborg2014Using}
Stine Lomborg and Anja Bechmann.
\newblock {Using APIs for data collection on social media}.
\newblock {\em The Information Society}, 30(4):256--265, 2014.

\bibitem{Oeldorf-Hirsch2020WhoFacebook}
Anne Oeldorf-Hirsch and Darren Gergle.
\newblock {“Who Knows What”: Audience Targeting for Question Asking on
  Facebook}.
\newblock {\em Proc. ACM Hum.-Comput. Interact.}, 4(GROUP), 1 2020.

\bibitem{MaiMonicaandLeung2020BigTwitter}
{Mai Monica and Leung, Carson K. and Choi Justin M C. and Kwan Long Kei
  Ronnie}.
\newblock {Big Data Analytics of Twitter Data and Its Application for Physician
  Assistants}.
\newblock In {Alhajj Reda and Moshirpour, Mohammad and Far Behrouz}, editor,
  {\em Data Management and Analysis: Case Studies in Education, Healthcare and
  Beyond}, pages 17--32. Springer International Publishing, Cham, 2020.

\bibitem{pfeffer2018tampering}
J{\"u}rgen Pfeffer, Katja Mayer, and Fred Morstatter.
\newblock Tampering with twitter’s sample api.
\newblock {\em EPJ Data Science}, 7(1):50, 2018.

\bibitem{Gjoka2010WalkingOSNs}
M~Gjoka, M~Kurant, C~T Butts, and A~Markopoulou.
\newblock {Walking in Facebook: A Case Study of Unbiased Sampling of OSNs}.
\newblock In {\em 2010 Proceedings IEEE INFOCOM}, pages 1--9, 2010.

\bibitem{Catanese2011CrawlingPurposes}
Salvatore~A Catanese, Pasquale De~Meo, Emilio Ferrara, Giacomo Fiumara, and
  Alessandro Provetti.
\newblock {Crawling Facebook for Social Network Analysis Purposes}.
\newblock In {\em Proceedings of the International Conference on Web
  Intelligence, Mining and Semantics}, WIMS ’11, New York, NY, USA, 2011.
  Association for Computing Machinery.

\bibitem{sirisuriya2015comparative}
De~S Sirisuriya et~al.
\newblock A comparative study on web scraping.
\newblock {\em Empty}, 2015.

\bibitem{Diouf2019WebApplication}
Rabiyatou Diouf, Edouard~Ngor Sarr, Ousmane Sall, Babiga Birregah, Mamadou
  Bousso, and Seny~Ndiaye Mbaye.
\newblock {Web Scraping: State-of-the-Art and Areas of Application}.
\newblock In {\em 2019 IEEE International Conference on Big Data (Big Data)},
  2019 IEEE International Conference on Big Data (Big Data), pages 6040--6042,
  Los Angeles, United States, 12 2019. IEEE.

\bibitem{Bergman2001}
Michael~K. Bergman.
\newblock White paper: The deep web: Surfacing hidden value.
\newblock {\em The Journal of Electronic Publishing}, 7(1), August 2001.

\bibitem{furche2013oxpath}
Tim Furche, Georg Gottlob, Giovanni Grasso, Christian Schallhart, and Andrew
  Sellers.
\newblock Oxpath: A language for scalable data extraction, automation, and
  crawling on the deep web.
\newblock {\em The VLDB Journal}, 22(1):47--72, 2013.

\bibitem{prifti2023emergence}
Ylli Prifti.
\newblock {\em The emergence of interpersonal and social trust in online
  interactions}.
\newblock PhD thesis, Birkbeck, University of London, 2023.

\bibitem{hursch1988sql}
Carolyn~J Hursch, Jack~L Hursch, and Carolyn~J Hursch.
\newblock {\em SQL, the Structured Query Language}.
\newblock Tab Books, 1988.

\bibitem{spertus2000squeal}
Ellen Spertus and Lynn~Andrea Stein.
\newblock Squeal: a structured query language for the web.
\newblock {\em Computer Networks}, 33(1-6):95--103, 2000.

\bibitem{lassila1998resource}
Ora Lassila, Ralph~R Swick, et~al.
\newblock Resource description framework (rdf) model and syntax specification.
\newblock {\em ""}, 1998.

\bibitem{mcbride2002four}
Brian McBride.
\newblock Four steps towards the widespread adoption of a semantic web.
\newblock In {\em International Semantic Web Conference}, pages 419--422.
  Springer, 2002.

\bibitem{Gunawan2019/03}
Rohmat Gunawan, Alam Rahmatulloh, Irfan Darmawan, and Firman Firdaus.
\newblock Comparison of web scraping techniques : Regular expression, html dom
  and xpath.
\newblock In {\em 2018 International Conference on Industrial Enterprise and
  System Engineering (ICoIESE 2018)}. Atlantis Press, 2019/03.

\bibitem{clark1999xml}
James Clark, Steve DeRose, et~al.
\newblock Xml path language (xpath), 1999.

\bibitem{Ehrenfeucht1974}
Andrzej Ehrenfeucht and Paul Zeiger.
\newblock Complexity measures for regular expressions.
\newblock In {\em Proceedings of the Sixth Annual ACM Symposium on Theory of
  Computing}, STOC ’74, page 75–79, New York, NY, USA, 1974. Association
  for Computing Machinery.

\bibitem{Sellers2011}
Andrew~Jon Sellers, Tim Furche, Georg Gottlob, Giovanni Grasso, and Christian
  Schallhart.
\newblock {OXPath}.
\newblock In {\em Proceedings of the 20th international conference companion on
  World wide web - {WWW} {\textquotesingle}11}. {ACM} Press, 2011.

\bibitem{furche2014diadem}
Tim Furche, Georg Gottlob, Giovanni Grasso, Xiaonan Guo, Giorgio Orsi,
  Christian Schallhart, and Cheng Wang.
\newblock Diadem: thousands of websites to a single database.
\newblock {\em Proceedings of the VLDB Endowment}, 7(14):1845--1856, 2014.

\bibitem{guo2019red}
Jinsong Guo, Valter Crescenzi, Tim Furche, Giovanni Grasso, and Georg Gottlob.
\newblock Red: Redundancy-driven data extraction from result pages?
\newblock In {\em The World Wide Web Conference}, pages 605--615, 2019.

\bibitem{fayzrakhmanov2018browserless}
Ruslan~R Fayzrakhmanov, Emanuel Sallinger, Ben Spencer, Tim Furche, and Georg
  Gottlob.
\newblock Browserless web data extraction: challenges and opportunities.
\newblock In {\em Proceedings of the 2018 World Wide Web Conference}, pages
  1095--1104, 2018.

\bibitem{hartig2018semantics}
Olaf Hartig and Jorge P{\'e}rez.
\newblock Semantics and complexity of graphql.
\newblock In {\em Proceedings of the 2018 World Wide Web Conference}, pages
  1155--1164, 2018.

\end{thebibliography}

\end{document}